\documentclass[pss]{wiley2sp} 
\usepackage{amsmath}

\newcommand{\system}{Sr$_3$Ru$_2$O$_7$}

\begin{document}

\title{Anisotropy of the low-temperature magnetostriction of \system}

\titlerunning{Anisotropy of the magnetostriction of \system}

\author{%
  C. Stingl\textsuperscript{\Ast,\textsf{\bfseries 1}},
  R.S. Perry\textsuperscript{\textsf{\bfseries 2}},
  Y. Maeno\textsuperscript{\textsf{\bfseries 3,4}},
   and
  P. Gegenwart\textsuperscript{\textsf{\bfseries 1}}}

\authorrunning{C. Stingl et al.}

\mail{e-mail
  \textsf{cstingl@gwdg.de}, Phone:
  +49-551-39-13142, Fax: +49-551-39-22328}

\institute{%
  \textsuperscript{1}\, I. Physikalisches Institut, Georg-August-Universit\"{a}t
G\"{oe}ttingen, Friedrich-Hund-Platz 1, 37077  G\"{oe}ttingen,
Germany\\
\textsuperscript{2}\, Scottish Universities Physics Alliance, School of Physics, University of Edinburgh, Mayfield Road, Edinburgh EH9 3JZ, Scotland\\
\textsuperscript{3}\, International Innovation Center, Kyoto University, Kyoto 606-8501, Japan\\
\textsuperscript{4}\, Department of Physics, Kyoto University, Kyoto 606-8502, Japan\\
    }

\received{XXXX, revised XXXX, accepted XXXX} 
\published{XXXX} 


\abstract{%
%
%
%
\abstcol{We use high-resolution capacitive dilatometry to study the
low-temperature linear magnetostriction of the bilayer ruthenate
Sr$_3$Ru$_2$O$_7$ as a function of magnetic field applied
perpendicular to the ruthenium-oxide planes ($B\parallel c$). The
relative length change $\Delta L(B)/L$ is detected either parallel
or perpendicular to the $c$-axis close to the metamagnetic region
near $B=8$~T.} {In both cases, clear peaks in the coefficient
$\lambda(B)=d(\Delta L/L)/dB$ at three subsequent metamagnetic
transitions are observed. For $\Delta L\perp c$, the third
transition at 8.1~T bifurcates at temperatures below 0.5~K. This is
ascribed to the effect of an in-plane uniaxial pressure of about
15~bar, unavoidable in the dilatometer, which breaks the original
fourfold in-plane symmetry.} }

\maketitle   

\section{Introduction}

The bilayer ruthenate Sr$_3$Ru$_2$O$_7$ has recently attracted much
interest, because of itinerant electron metamagnetism and the
possible formation of a nematic electron fluid close a quantum
critical point near 8~T for fields applied perpendicular to the
ruthenium-oxide planes, i.e. parallel to the tetragonal
c-axis~\cite{Grigera01,Grigera03,Grigera04,Chufo}. Early studies of
the magnetic susceptibility of single crystals have suggested
quantum criticality to arise from the suppression of the critical
temperature of a first-order metamagnetic transition by tuning the
field angle towards $B\parallel c$ \cite{Grigera03}. Subsequent
studies on high-quality single crystals ($\rho_{0}=0.4\mu\Omega$cm)
have revealed a fine structure in the $T$-$B$ phase diagram, bound
by two metamagnetic transitions at 7.85 and 8.07~T which are of
first-order for temperatures below 0.7 and 0.5~K,
respectively~\cite{Grigera04,Weickert}. Within this regime, the
electrical resistivity peaks and becomes temperature independent,
indicating an increase of elastic scattering, possibly due to the
formation of some kind of domains~\cite{Grigera04}, whereas outside
this region the thermal expansion behavior was found to be
compatible with metamagnetic quantum criticality~\cite{Gegenwart}.
An in-plane anisotropy of the electrical resistivity arises when the
applied magnetic field is tilted by $13^\circ$~off the $c$-axis,
indicating a spontaneously broken fourfold rotational symmetry in
the ab plane perpendicular to the c-axis~\cite{Chufo}. The coupling
of this presumed "electronic nematic state" to the lattice could be
studied most sensitively by capacitive dilatometry.
Previously, a strong magnetoelastic coupling with
highly enhanced magnetic Gr\"uneisen parameter has been found in
linear magnetostriction measurements along the c-axis in \system
\cite{Grigera04,GegenwartPhysica06}. Since in the novel phase the
four-fold symmetry is broken, length measurements perpendicular to
the c-axis are of particular interest. In this paper, we compare for
$B\parallel c$ the magnetostriction along and perpendicular to the
c-axis.

\section{Experiment}

For our experiments, we have used one piece of about (1.5~mm)$^3$
dimension of the same high-quality single crystal, grown by floating
zone technique~\cite{Perry_PRL_04}, which has been studied
previously in thermal expansion and magnetostriction along the
$c$-axis~\cite{Grigera04,Gegenwart,GegenwartPhysica06} (the original
crystal has broken in two pieces).\\The magnetostriction has been
determined by a miniaturized capacitive dilatometer, which is small
enough to be mounted in parallel and perpendicular configuration in
a dilution refrigerator with 18~T superconducting magnet. For our
measurements, we have applied the field parallel to the
$c$-direction, similar as previously. The linear magnetostriction
has been detected in two separate runs first parallel to the applied
field, i.e. $\Delta L\parallel B \parallel c$ as sketched in Figure
1a, and subsequently perpendicular to the field, $\Delta L\perp
B\parallel c$ (Fig. 2a)\footnote{The sample edges are parallel to
the axes of the pseudotetragonal crystal structure with $a\approx
b\approx3.89$\,\AA~\cite{kiyanagi_jpsj_04}.}.  The magnetic field is
varied with a rate of 1\,T/h for temperatures between 30\,mK and
4\,K. No hysteresis larger than 2~mT could be detected similar as
previously~\cite{GegenwartPhysica06}. The magnetostriction
coefficient $\lambda=d(\Delta L/L)/dB$ has been obtained by linear
fits on 20~mT field intervals.

\section{Results}

\begin{figure}[t]
\includegraphics*[width=\linewidth]{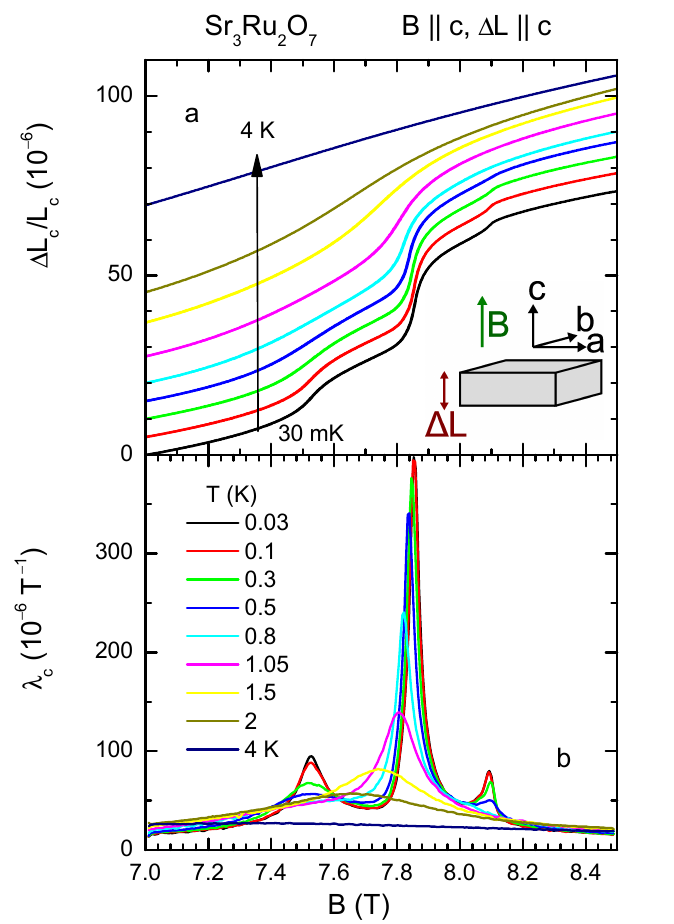}
\caption{a: Linear magnetostriction $\Delta L_c/L_c$ along the
$c$-axis of Sr$_3$Ru$_2$O$_7$ measured at various temperatures for
$B\parallel c$, as indicated in the sketch. b: Respective
magnetostriction coefficient $\lambda_c(B)$.} \label{lambda_c}
\end{figure}

\begin{figure}[t]
\includegraphics*[width=\linewidth]{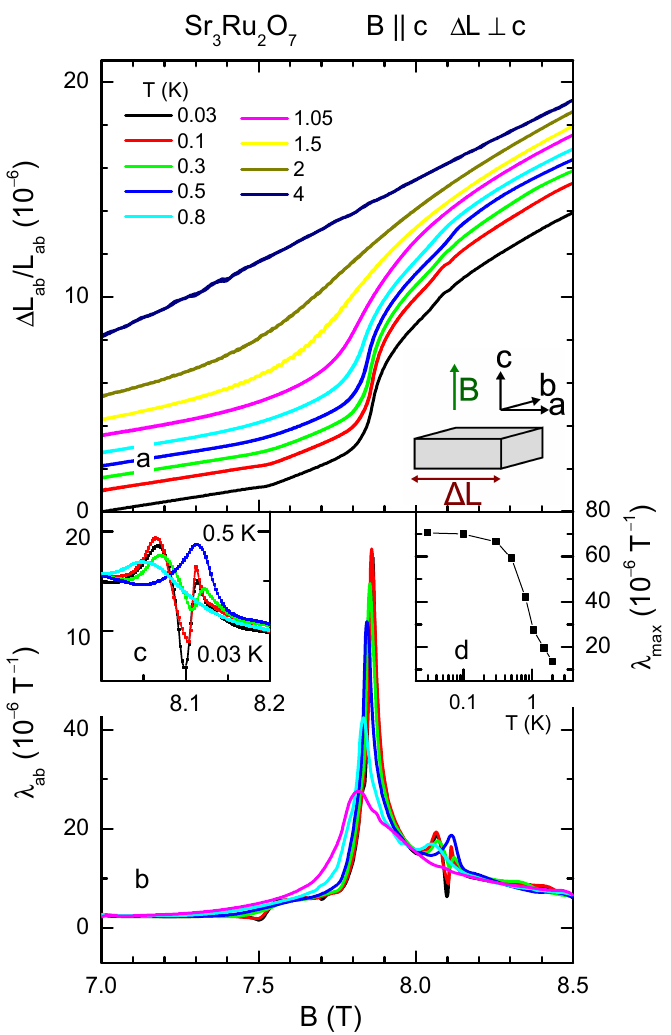}
\caption{a: Linear magnetostriction $\Delta L_{ab}/L_{ab}$ of
Sr$_3$Ru$_2$O$_7$ measured perpendicular to the $c$-axis at various
temperatures for $B\parallel c$, as indicated in the sketch. b:
Respective magnetostriction coefficient $\lambda_{ab}(B)$. Inset c
enlarges region close to upper metamagnetic transition. Inset d
displays temperature dependence (on a log scale) of maxima in
$\lambda_{ab}$ at the central metamagnetic transition.}
\label{lambda_ab}
\end{figure}

The length change parallel to the field and to the $c$-axis, $\Delta
L_c/L_c$ between 7 and 8.5~T, displayed in Figure~\ref{lambda_c}a,
is similar as observed
previously~\cite{Grigera04,GegenwartPhysica06}. However, the three
peaks in the coefficient $\lambda_c(B)$ are narrower and larger,
indicating sharper transitions in this piece of the original single
crystal used in \cite{Grigera04,GegenwartPhysica06}. This may be
related to the first-order nature of the metamagnetic transitions.
The height of $\lambda_c$ at the central peak saturates at low-$T$,
similar as found previously~\cite{GegenwartPhysica06} and similar as
observed for $\lambda_{ab}$ (inset d of Fig.~2).

Figure \ref{lambda_ab} shows corresponding magnetostriction results
transverse to the applied field. In this configuration, the length
change $\Delta L_{ab}/L_{ab}$ between 7 and 8.5~T is about five
times smaller compared to the former case. The central peak in
$\lambda_{ab}$ is qualitatively similar as found for $\lambda_c$,
including its temperature dependence. However, the two other
metamagnetic transitions display very different signatures. The
7.5~T crossover is visible as a distinct {\it minimum} in
$\lambda_{ab}(B)$ at lowest temperatures. Furthermore, a splitting
of the 8.1~T transition is observed below 0.5~K, resulting in a
sharp minimum close to 8.1~T in $\lambda_{ab}(B)$ at lowest
temperatures (cf. inset c of Fig.~2). As discussed below, we ascribe
this anomaly to the small uniaxial pressure generated by our
dilatometer.

\begin{figure}[t]
\includegraphics*[width=\linewidth]{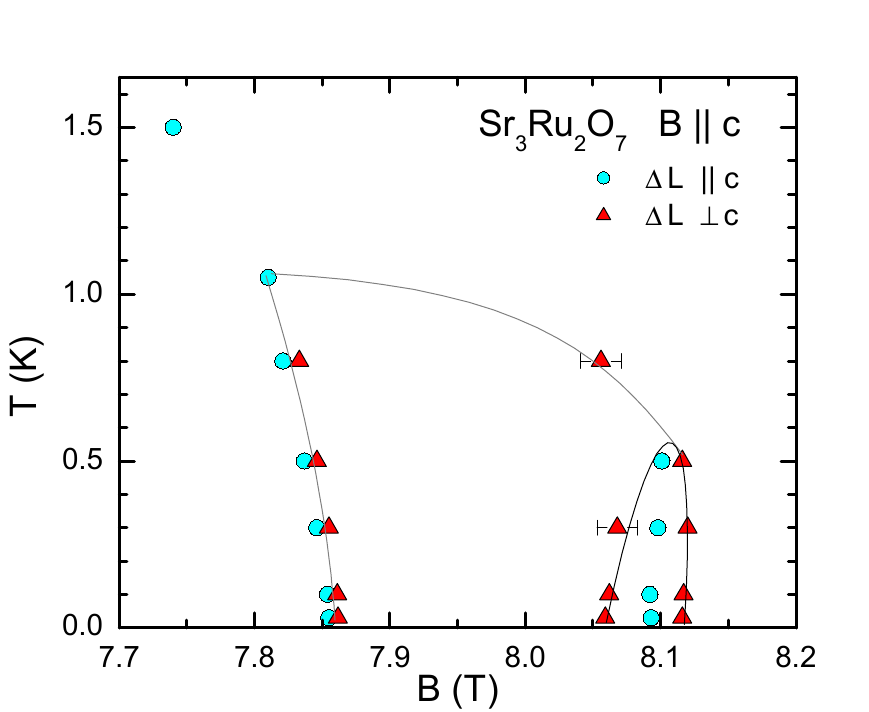}
\caption{$T$-$B$ phase diagram for \system, $B\parallel c$. Blue
circles and red triangles indicate positions of maxima in
$\lambda_c$ and maxima and minima in $\lambda_{ab}$. Black and gray
lines indicate bifurcation of 8.1~T transition in $\lambda_{ab}$
measurement and boundary of proposed symmetry-broken
phase~\cite{Grigera04}, respectively.} \label{phasediagram}
\end{figure}

Figure 3 displays a $T$-$B$ phase diagram with the positions of
maxima in the linear magnetostriction coefficients $\lambda_{ab}(B)$
and $\lambda_c(B)$ indicated by blue circles and red triangles,
respectively. Besides the bifurcation of the 8.1~T peak for
measurements perpendicular to the c-axis at temperatures below
0.5~K, we also note the small shift of the central peak towards
larger $B$ for this configuration.

\section{Discussion}

The sign of the linear magnetostriction is related by the Maxwell
relation $\lambda V_\text{m}=-(dM/dP)_{P\rightarrow 0}$ to the uniaxial
pressure dependence of the magnetization ($V_\text{m}$: molar volume).
Since $\lambda>0$ for both orientations for the main metamagnetic
transition, the magnetization decreases with increasing pressure, in
qualitative agreement with the shift of the metamagnetic field under
hydrostatic pressure~\cite{Chiao}.

Positions of transitions in $T$-$B$ phase space can generally not
depend on the direction along which magnetostriction has been
measured (for similar field orientation). Since the {\it same}
crystal has been used, the only reason for the observed differences
in the positions of the metamagnetic transitions is the effect of a
weak uniaxial pressure in these measurements. The two parallel flat
springs of our dilatometer exert a force of approximately 3~N on the
samples cross-section along the measurement direction, corresponding
to a uniaxial pressure of roughly 15~bar for the studied crystal.
Using the estimated uniaxial pressure dependence of the central
metamagnetic transition~\cite{GegenwartPhysica06}, this pressure
causes a shift of approximately 0.01~T, only, which is of similar
size as a possible field offset due to remanence in the
superconducting magnet. However, for the $\lambda_{ab}$ measurements
the uniaxial pressure acts {\it perpendicular} to the $c$-axis and
therefore breaks the fourfold in-plane symmetry. Such symmetry
breaking could have a strong effect on the low-$T$ properties.
Most interestingly, the 8.1~T transition is found to bifurcate below
0.5~K, i.e. at temperatures below which this metamagnetic transition
is of first order \cite{Grigera04}. The comparison with the
magnetostriction experiments along the c-axis indicates that this
splitting arises from the in-plane uniaxial pressure.
Interestingly, a bifurcation of two metamagnetic transitions has
also been found for a high-quality Sr$_3$Ru$_2$O$_7$ single crystal
at $B\perp c$, for temperatures below 0.5~K~\cite{Perry_JPSJ_05}. In this
case, the field applied perpendicular to the $c$-axis breaks acts
symmetry breaking.

To summarize, we have studied the anisotropy of the low-temperature
magnetostriction of Sr$_3$Ru$_2$O$_7$ by capacitive measurements
along and perpendicular to the applied field $B\parallel c$. $\Delta
L_c/L_c(B)$ is about five times larger than $\Delta
L_{ab}/L_{ab}(B)$. Remarkably, we observe a splitting of the
metamagnetic transition at 8.1~T for the measurement perpendicular
to the $c$-axis that is ascribed to a symmetry breaking uniaxial
pressure of about 15 bar in this experiment.

We thank R. K\"{u}chler for his help in the construction of the
miniaturized dilatometer. This work was supported by the Deutsche
Forschungsgemeinschaft within SFB~602.


\end{document}